\documentclass{nature}
\usepackage{graphicx,color,epsfig,times}

\usepackage{graphicx,color,epsfig}
\title{Direct measurement of antiferromagnetic domain fluctuations}

\author{O. G. Shpyrko$^{1}$, E. D. Isaacs$^{1,2}$, J. M. Logan$^{2}$, Yejun
Feng$^{2}$, G. Aeppli$^{3}$, R. Jaramillo$^{2}$, H. C. Kim$^{2}$, T. F.
Rosenbaum$^{2}$, P. Zschack$^{4}$, M. Sprung$^{4}$, S. Narayanan$^{4}$ and
A. R. Sandy$^{4}$}

\begin{document}

\maketitle

\begin{affiliations}
\item  \textit{Center for Nanoscale Materials, Argonne National Laboratory, Argonne, IL 60439, USA}
\item  \textit{James Franck Institute and Dept. of Physics, University of Chicago, Chicago, IL 60637, USA}
\item \textit{London Centre for Nanotechnology and Department of Physics and Astronomy, University College London, London WC1E 6BT, UK}
\item \textit{Advanced Photon Source, Argonne National Laboratory, Argonne, IL 60439, USA}
\end{affiliations}

\begin{abstract}Measurements of magnetic noise emanating from ferromagnets due
to domain motion were first carried out nearly 100 years ago [1] and have underpinned much science and technology [2, 3].
Antiferromagnets, which carry no net external magnetic dipole moment, yet
have a periodic arrangement of the electron spins extending over macroscopic
distances, should also display magnetic noise, but this must be sampled at
spatial wavelengths of order several interatomic spacings, rather than the
macroscopic scales characteristic of ferromagnets. Here we present the first
direct measurement of the fluctuations in the nanometre-scale spin-
(charge-) density wave superstructure associated with antiferromagnetism in
elemental Chromium. The technique used is X-ray Photon Correlation
Spectroscopy, where coherent x-ray diffraction produces a
\textit{speckle}  pattern that serves as a ``fingerprint''
of a particular magnetic domain configuration. The temporal evolution of the
patterns corresponds to domain walls advancing and retreating over micron
distances. While the domain wall motion is thermally activated at
temperatures above 100K, it is not so at lower temperatures, and indeed has
a rate which saturates at a finite value -- consistent with quantum
fluctuations - on cooling below 40K. Our work is important because it
provides an important new measurement tool for antiferromagnetic domain
engineering as well as revealing a fundamental new fact about spin dynamics
in the simplest antiferromagnet.
\end{abstract}

  Because of scientific and technical interest in ferromagnetic domains,
there has been large, long-standing activity on magnetic noise in
ferromagnets as a direct witness of domain motion. As antiferromagnets begin
to find applications themselves, for example as pinning layers in
spintronics, there is a need for measurements of the noise associated with
moving antiferromagnetic domains. Antiferromagnetic domain dynamics are also
important because they are implicated in basic problems in condensed matter
physics, such as high temperature superconductivity and `heavy' Fermions.
Neutrons are an excellent non-local probe of antiferromagnetism and its
dynamics [4]. However, a direct \textit{local} probe of mesoscopic antiferromagnetic
domain dynamics has not been hitherto available because the magnetic dipole
moments for antiferromagnets vanish on the scale of a nanometer, rendering
the domain fluctuations responsible for noise essentially invisible to the
direct magnetometer probes (e.g. superconducting interference devices) which
have been so successful for ferromagnets [5].

  Chromium is a body-centred cubic (bcc) metal with an antiferromagnetic
state nearly described by the simple rule that the electrons surrounding
each Cr atom have magnetization opposite to those on the nearest neighbour
Cr atoms. What actually occurs is sinusoidal modulation of this elementary
magnetic structure, called a spin density wave (SDW) with wavelength
$\lambda $=6-8 nm, along one of the three equivalent cubic (100) directions.
A single crystal chromium sample cooled below the N\'{e}el temperature
T$_{N}$=311 K spontaneously breaks (see Fig. 1) into three types of magnetic
domains characterized by the three different choices for the SDW propagation
direction [6]. The SDW is accompanied by a charge density wave (CDW), a
combination of both itinerant and ionic charge modulation.

  X-ray microdiffraction reveals that the typical size of the SDW domains
in bulk Cr samples is on the order of 1-30 $\mu $m [7]. Fluctuations of
domain walls at fixed temperature have been studied via random electrical
telegraph noise in thin Cr films for temperatures above 140 K [8]. Even
though the measurements were done for mesoscopic samples, the effects on the
electrical resistance R of the switching dynamics were small ($\delta
$R/R$\sim $10$^{ - 5})$ and the interpretation difficult because R is an
indirect probe of the underlying SDW and CDW order.

  We report the first direct observations of domain wall fluctuations in
bulk Cr using X-ray Photon Correlation Spectroscopy (XPCS), which overcomes
the limitations of the classic bulk and laser probes in that it accesses the
short wavelength structure associated with the SDW directly. A coherent beam
illuminating a partially ordered system (in our case consisting of SDW/CDW
domains) produces an interference pattern, also known as \textit{speckle} [9, 10]. Due to
the high sensitivity of speckle to minute changes in domain wall
configuration, the time variation of the speckle pattern directly reveals
the dynamics of domain structure. Fig. 2a is a schematic of the experimental
configuration, and Figure 2b shows a speckle pattern of the (200) Bragg peak
for the bcc Cr lattice. Interference fringes arising from partial coherence
of the x-ray beam are clearly seen in the image as well as in the line scans
shown in Fig. 2c. Incoherent diffraction would produce the Gaussian-like
profile represented by the black line in Fig. 2c. The lattice Bragg speckle
pattern is static over 5 hrs indicating the high level of stability for our
instrumentation and the sample.

  We turn next to the speckle pattern for the (2-2$\delta $, 0, 0) CDW
superlattice reflection, displayed for 17 K at a variety of times in Fig.
3b. The patterns in subsequent frames, separated by 1,000 s, grow
increasingly dissimilar for longer time lags -- patterns within frames
collected more than 3,000 s apart appear completely uncorrelated. Thus, the
CDW speckle evolves with a characteristic time of a few thousand seconds or
less, much shorter than the $>$20,000 s relaxation time for the bcc Bragg
speckle of Fig. 2c. This indicates that the changes in the CDW speckle are
indeed due to changes in the magnetic domain configuration, rather than some
experimental artefact. For example, drift of the x-ray beam or the cryostat,
motion of crystalline defects within the Cr sample or any other effect not
related to magnetic domain dynamics would inevitably cause changes in\textit{ both }the CDW
and (200) Bragg speckle.

  The spatial sensitivity of the speckle to domain motion is described by
two distinct lengths: the first is 1/$\Delta $Q$\sim $100{\AA}, where
$\Delta $Q=10$^{ - 2}${\AA}$^{ - 1}$ is the total size of visible speckle
pattern in reciprocal space (See Figs. 2b, 2c and 3b) and represents the
minimum size of domains with a visible impact on the speckle pattern. The
second is the domain wall displacement necessary to produce a speckle
pattern that is highly dissimilar (or uncorrelated) to the original one. A
combination of x-ray microdiffraction images of domain configurations and
speckle simulations indicate that this second length is 1 $\mu $m (see
Methods and Supplementary Information).

  Beyond revealing that domain walls are moving by distances of order 1
$\mu $m, the data provide several other important quantities. For example,
we can evaluate the autocorrelation function g$_{2}$(t):
\begin{equation}
\label{eq1}
g_2 (t) = \frac{\left\langle {I(\tau )I(\tau + t)} \right\rangle _\tau
}{\left\langle {I(\tau )} \right\rangle _\tau ^2} = 1 + A\left| {F(Q,t)}
\right|^2
\end{equation}
where I($\tau$) and I($\tau+t$) are the intensities
in a given pixel for frames taken at times $\tau $ and $\tau $+t
respectively, F(Q,t) is the intermediate scattering function, $A$ describes
the beam coherence [9,10], and the averaging is performed over times $\tau $
and pixels. Figure 3a shows $\vert $F(Q,t)$\vert ^{2 }$for several
temperatures calculated from the CDW speckle. For large time delays the
speckle patterns become uncorrelated, resulting in g$_{2}$(t)=1,
corresponding to $\vert $F(Q,t)$\vert ^{2}$ =0. The dynamics are strongly
temperature-dependent: upon cooling, the domain fluctuation times increases
by nearly two decades. Surprisingly, below 40 K the times remain finite,
rather than diverging as expected for thermally driven fluctuations.

  Two distinct fluctuation timescales are visible in most datasets
presented in Fig. 3a. The measured $\vert $F(Q,t)$\vert $ was therefore
modelled by a double exponential form:
\begin{equation}
\label{eq2}
| F(Q,t)|= a \exp[- (t /\tau_{F})^\beta] + (1 - a) \exp [- (t /\tau _{S})^\beta ]
\end{equation}
A small value of a=0.03-0.10 indicates that the time dependence of $\vert
$F(Q,t)$\vert $ is mainly due to slow fluctuations. The value of the
stretching exponent $\beta $ was found to be greater than 1, manifested by
the ``compressed'' shape of the $\vert $F(Q,t)$\vert $. Compressed
exponential relaxation has been observed for a variety of soft matter
systems undergoing ``jamming'' transitions which results in arrested,
solid-like collective dynamics [11, 12, 13] with $\beta >$1, as instead of
liquid-like fluctuations with $\beta  \le $1. Extended to our system, this
points to elastically coupled dynamics between blocks of spins, similar to
elastic collective depinning dynamics observed in CDW conductors [14], an
observation also consistent with the weakly pinned nature of SDW/CDW domains
[15-17]. Furthermore, the fit value of $\beta $ at T$<$100K is approximately
1.5 (Fig. 4b), a universal value for dynamics of soft condensed matter
systems in a jammed state [18].

Fig. 4 shows the T-dependence of the slow relaxation times $\tau _{S}$
obtained from fits to autocorrelation functions in Fig. 3a. The 20{\%}
uncertainty in fitting parameters $\tau _{S }$arises primarily from
counting statistics of the autocorrelation function g$_{2}$(t) (see
Supplementary Information). Standard thermal activation ($\tau_t^{-1}=f_{o} \exp(-\Delta E/k_{B}T)$, blue line) with a single
attempt frequency $f_{o}$ and activation barrier $\Delta E/k_{B }=240\pm50 K$ accounts for the data at high T. The thermal picture fails
spectacularly at low temperature for T$<$40 K, and a switching mechanism which
is temperature-independent in this range is required. The simplest
possibility is that switching between low-energy domain wall configurations
occurs via quantum tunneling, rather than classical thermal activation. A
fit to the data that combines a thermally activated model and a quantum
tunneling contribution represented by a temperature-independent residence
time $\tau _{QT}$, $\tau _{S}^{-1}=\tau_{QT}^{-1}+\tau _{R}^{-1}\exp(-\Delta E/k_{B}T)$ is shown by the red solid line in Fig. 4 for $\tau
_{QT}$=5,000~s and $\tau _{R}=$15~s (confidence
limits obtained from the fits are $\tau _{QT}$=5,000$\pm $1,000~s and
t$_{R}$=4-60~s). The short-term fluctuation rate $\tau _{F}^{-1}$ observed in the autocorrelation data in Fig. 3a has the same magnitude
as the attempt frequency $\tau _{R}^{-1}$. In analogy with
alpha and beta relaxation observed in glasses, supercooled liquids and
jammed soft matter systems, faster fluctuations represent local relaxation,
while slower fluctuations are due to collective relaxation modes.

  The relaxation times observed here are similar to those associated with
magnetization switching in ferromagnets first observed by Barkhausen [1] and
studied since then in systems from bulk materials to magnetic molecules [5,
19-22]. Antiferromagnets have more complex order than ferromagnets because
they break translation as well as spin rotation invariance, which has forced
us to formulate a very crude physical picture, to understand our data at a
semi quantitative level. We start with the realization that to minimize the
very large exchange energy ($>$0.4~eV) [23, 24] associated with domain walls,
it is clearly advantageous for the nodal planes (where the spin polarization
vanishes) of the SDW with its propagation vector perpendicular to the domain
wall to lie on the walls [25]. Such an assumption is further supported by
the previously observed preference for the formation of SDW nodes at Fe/Cr
interfaces [26]. This implies that the fundamental switching unit (grey
shaded region in Fig. 1a) is of volume $V_{S}\sim(\lambda/2)^{3}$
where $\lambda$ is the underlying period of the SDW. In the simplest
Gaussian model where underlying units are switching randomly at typical
times $\tau _{U}$, we would conclude that the switching time for a volume
of V=1 $\mu $m$^{3}$ would be (V/V$_{S})^{2}\tau _{U}$. Using our
experimental value for the attempt frequency $\tau_{R}^{-1}$, we
therefore obtain $\tau _{U}^{-1}$= 36~THz=140~meV as the attempt
frequency for rotating an entire unit. This is an electronic energy scale,
and could therefore be derived from the hopping of electrons across the
domain wall; such electrons (also important in electrical noise measurements
[8]) are, after all, responsible for the current fluctuations which sample
the possibility of rotating the Fermi surface of a `quantum dot' with the
fundamental unit volume V$_{S}$. It is fortunate that the barrier for
rotation between two minima has been identified by neutron spectroscopy on
Cr$_{0.95}$V$_{0.05}$ (data for pure Cr are not published) (See Fig. 2 of
Ref. 27) as the energy E$_{o}$ at which the incommensurate spin density
fluctuations no longer display distinct peaks at the incommensurate
satellite positions; E$_{o}$ is found to be of order 25~meV (or 290~K),
which is not far from the tunnelling barrier $\Delta $E=20~meV (or 240~K)
established in our own experiments. Interestingly, it is this energy, rather
than the much larger exchange coupling, which corresponds to T$_{N}$.

  In the simplest WKB approximation (see e.g. Ref. 22), the dimensionless
ratio $\tau _{R}$/$\tau _{QT}$is equal to $\exp (-S/\hbar)$,
where $S = \sqrt {I\Delta E} $ is the tunneling action. Because the
underlying attempt frequencies and their rescaling to account for observable
effects in the X-ray experiment are the same for both incoherent quantum and
classical processes, all of the detail -- invoking multiple rotors - of the
last paragraph drops out, and S characterizes a single rotor. We can
therefore calculate the moment of inertia I of the quantum rotor using the
measured parameters $\tau _{QT}$, $\tau _{R}$ and the barrier height
$\Delta $E=20 meV obtained from the Arrhenius regime. The result is I=100
m$_{e}$ nm$^{2}$, which, assuming a cube of uniform density distributed over
the ($\lambda $/2)$^{3}$ volume of the fundamental unit corresponds to 0.1
electron mass m$_{e}$ per Chromium unit cell. This remarkable result,
derived only from our data and the simple physical picture of Fig. 1, is
consistent with Hall effect data [28, 29] showing that the SDW is associated
with the loss of a similar number of carriers, which of course must be moved
with the rotors when there is a switching event.

  We have introduced the direct measurement of noise spectra in
\textit{antiferromagnets}. Our experiments access local mesoscale spin dynamics with just a few
domain walls in the illuminated volume, an advantage over non-local
experimental probes that cannot be easily applied for macroscopic or bulk
structures. The key finding is that even in bulk samples, and at
temperatures very low compared to the Neel temperature, domain walls can be
unstable on time scales of fractions of an hour. What this means is that the
stability of antiferromagnetism needs to engineered, e.g. by insertion of
appropriate pinning centres, into devices that exploit it. This will become
even more important for nanoscale spintronics including antiferromagnetic
elements. Beyond the obvious advantages for magnetic engineering of now
having a technique with which antiferromagnetic domain fluctuations can be
readily assessed, we foresee tremendous opportunities in areas such as the
science of antiferromagnetic nanoparticles.

\textbf{Methods.} Experiments were carried out at beamlines 33-ID
and 8-ID of Advanced Photon Source, Argonne National Laboratory. The
undulator-generated x rays are monochromatized by a Si (111) crystal at an
energy E=7.35 keV (wavelength 1.686 Angstroms). A 10 $\mu $m pinhole
aperture or a 10${\rm g}\mu $m (horizontally) by 40 $\mu $m (vertically)
slits placed 5 cm upstream from the sample selected partially coherent
portion of the x-ray beam with a resulting coherence fraction A$ \approx
$0.07-0.18. A high purity (111) Cr wafer (Alfa Aesar, Ward Hill, MA) was
used to ensure roughly equal population of domains. The sample was mounted
inside a low-drift He flow cryostat, with thermal shielding provided by 600
$\mu $m thick Be dome. Speckle patterns were recorded with a Princeton
Instruments PI-LCX 1300 deep depletion x-ray CCD camera (1340x1300 pixel
array with 20 micron by 20 micron pixel size), located 150 cm from the
sample in reflection geometry.

\textbf{References.}
 \newline
 [1] Barkhausen, H. Zwei mit Hilfe der Neuen Verst\"{a}rker
entdeckte Erscheinungen.\textit{ Phys. Z. }\textbf{20}, 401-403 (1919).
\newline
 [2] Weissman, M. B. Low Frequency Noise as a tool to study
disordered materials. \textit{Annu. Rev. Mater. Sci.} \textbf{26}, 395-429 (1996).
\newline
 [3] Sethna, J. P., Dahmen, K. A. {\&} Myers C. R. Crackling Noise.
\textit{Nature} \textbf{410}, 242-250 (2001).
\newline
 [4] Ekspong, G. (ed.) \textit{Nobel Lectures, Physics 1991-1995} (World Scientific Publishing Co., Singapore,
1997). 1994 Nobel prize lectures by Clifford G. Shull and Bertram N.
Brockhouse (pp. 107-154)
\newline
 [5] Vitale, S., Cavalieri, A., Cerdonio, M., Maraner, A. {\&}
Prodi, G. A. Thermal equilibrium noise with 1/f spectrum in a ferromagnetic
alloy: Anomalous temperature dependence. \textit{J. Appl. Phys. }\textbf{76},
6332-4 (1994).
\newline
 [6] For a review of SDW in Cr see Fawcett, E. Spin-density-wave
antiferromagnetism in chromium. \textit{Rev. Mod. Phys.} \textbf{60}, 209-283 (1988).
\newline
 [7] Evans, P.G., Isaacs, E.D., Aeppli, G., Cai, Z.-H. {\&} Lai, B.
X-ray microdiffraction image of antiferromagnetic domain evolution in
chromium. \textit{Science} \textbf{295}, 1042-1045 (2002).
\newline
 [8] Michel, R. P., Israeloff, N. E., Weissman, M. B., Dura, J. A.
{\&} Flynn, C. P. Electrical-noise measurements on chromium films. \textit{Phys. Rev. B }\textbf{44}, 7413-7425 (1991).
\newline
 [9] Sutton, M., Mochrie, S. G. J., Greytak, T., Nagler, S. E. {\&}
Berman, L. E. Observation of speckle by diffraction with coherent X-rays.
\textit{Nature} \textbf{352}, 608-610 (1991).
\newline
 [10] Sutton, M. Coherent X-ray Diffraction. In Mills, D. (ed.)
\textit{Third-Generation Hard X-Ray Synchrotron Radiation Sources: Source Properties, Optics, and Experimental Techniques} (John Wiley {\&} Sons, New York, 2002).
\newline
 [11] Cipelletti, L., Manley, S., Ball, R. C. {\&} Weitz, D. A.
Universal Aging Features in the Restructuring of Fractal Colloidal Gels.
\textit{Phys. Rev. Lett.} \textbf{84}, 2275--2278 (2000).
\newline
 [12] Bandyopadhyay, R. \textit{et al.} Evolution of Particle-Scale Dynamics in an
Aging Clay Suspension. \textit{Phys. Rev. Lett.} \textbf{93}, 228302 (2004).
\newline
 [13] Falus, P., Borthwick, M. A., Narayanan, S., Sandy, A. R. {\&}
Mochrie, S. G. J. Crossover from Stretched to Compressed Exponential
Relaxations in a Polymer-Based Sponge Phase. \textit{Phys. Rev. Lett.} \textbf{97}, 066102 (2006).
\newline
 [14] Lemay, S. G., Thorne, R. E., Li Y. {\&} Brock, J. D.
Temporally ordered collective creep and dynamic transition in the
charge-density-wave conductor NbSe3. \textit{Phys. Rev. Lett.} \textbf{83}, 2793-2796 (1999).
\newline
 [15] Fukuyama H.$^{ }${\&} Lee, P. A. Dynamics of the
charge-density wave. I. Impurity pinning in a single chain. \textit{Phys. Rev. B} \textbf{17}, 535 - 541 (1978).
\newline
 [16] Fukuyama H.$^{ }${\&} Lee, P. A. Dynamics of the
charge-density wave. II. Long-range Coulomb effects in an array of chains.
\textit{Phys. Rev. B} \textbf{17}, 542 - 548 (1978).
\newline
 [17] Littlewood, P. B. {\&} Rice, T. M. Metastability of the $Q$
Vector of Pinned Charge- and Spin-Density Waves. \textit{Phys. Rev. Lett.}\textbf{\textit{}}\textbf{48}, 44 - 47 (1982).
\newline
 [18] Cipelletti, L. \textit{et al.} Universal non-diffusive slow dynamics in aging
soft matter. \textit{Faraday Discuss.} \textbf{123}, 237-251 (2003).
\newline
 [19] Chudnovsky, E. M. {\&} Tejada, J. \textit{Macroscopic Quantum Tunneling of the Magnetic Moment} (Cambridge University Press, Cambridge, UK, 1998).
\newline
 [20] Barbara, B.\textit{ et al.} Quantum tunneling in magnetic systems of various
sizes.\textit{ J. Appl. Phys.} \textbf{73}, 6703-6706 (1993).
\newline
 [21] Wernsdorfer, W. Classical and quantum magnetization reversal
studied in nanometersized particles and clusters.\textit{ Adv. Chem. Phys.} \textbf{118}, 99 (2001).
\newline
 [22] Brooke, J., Rosenbaum, T. F. {\&} Aeppli, G. Tunable quantum
tunnelling of magnetic domain walls. \textit{Nature}, \textbf{413}, 610-613 (2001).
\newline
 [23] Fenton, E. W. {\&} Leavens, C. R. The spin density wave in
chromium. \textit{J. Phys. F} \textbf{10}, 1853-1878 (1980).
\newline
 [24] Fenton, E. W. Domains in the Spin-Density-Wave Phases of
Chromium.\textit{ Phys. Rev. Lett.} \textbf{45}, 736 - 739 (1980).
\newline
 [25] Michel, R. P., Weissman, M. B., Ritley, K., Huang, J. C. {\&}
Flynn, C. P. Suppression of polarization fluctuations in chromium alloys
with commensurate spin-density waves. \textit{Phys. Rev. B} \textbf{47}, 3442 - 3445 (1993).
\newline
 [26] Fullerton, E. E., Bader, S. D. {\&} Robertson, J. L.,
Spin-density-wave Antiferro-magnetism of Cr in Fe/Cr(001) Superlattices,\textit{ Phys. Rev. Lett.} \textbf{77}, 1382-1385 (1996).
\newline
 [27] Hayden, S. M., Doubble, R., Aeppli, G., Perring, T. G. {\&}
Fawcett, E. Strongly Enhanced Magnetic Excitations Near the Quantum Critical
Point of Cr$_{1 - x}$V$_{x}$ and Why Strong Exchange Enhancement Need Not
Imply Heavy Fermion Behavior. \textit{Phys. Rev. Lett.} \textbf{84}, 999 - 1002 (2000).\newline
 [28] Lee, M., Husmann, A., Rosenbaum, T. F. {\&} Aeppli, G. High
resolution study of magnetic ordering at absolute zero. \textit{Phys. Rev. Lett.} \textbf{92}, 187201 (2004).
\newline
 [29] Yeh, A. \textit{et al.} Quantum phase transition in a common metal. \textit{Nature} \textbf{419}, 459-462 (2002).
\newline

\textbf{Supplementary Information} accompanies the paper.
\begin{addendum}
 \item Use of the Center for Nanoscale Materials
and Advanced Photon Source was supported by the U. S. Department of Energy,
Office of Science, Office of Basic Energy Sciences. The work at the
University of Chicago was supported by the National Science Foundation,
while that in London was funded by a Royal Society Wolfson Research Merit
Award and the Basic Technologies programme of RCUK. Authors declare they
have no competing financial interests.
\item[Competing Interests] The authors declare that they have no
competing financial interests.
\item[Correspondence] Correspondence and requests for materials should be addressed to O.~G.~S. (oshpyrko@anl.gov)
\end{addendum}

\newpage

\begin{figure}
\vspace{5mm}
\includegraphics[angle=0,width=0.7\columnwidth]{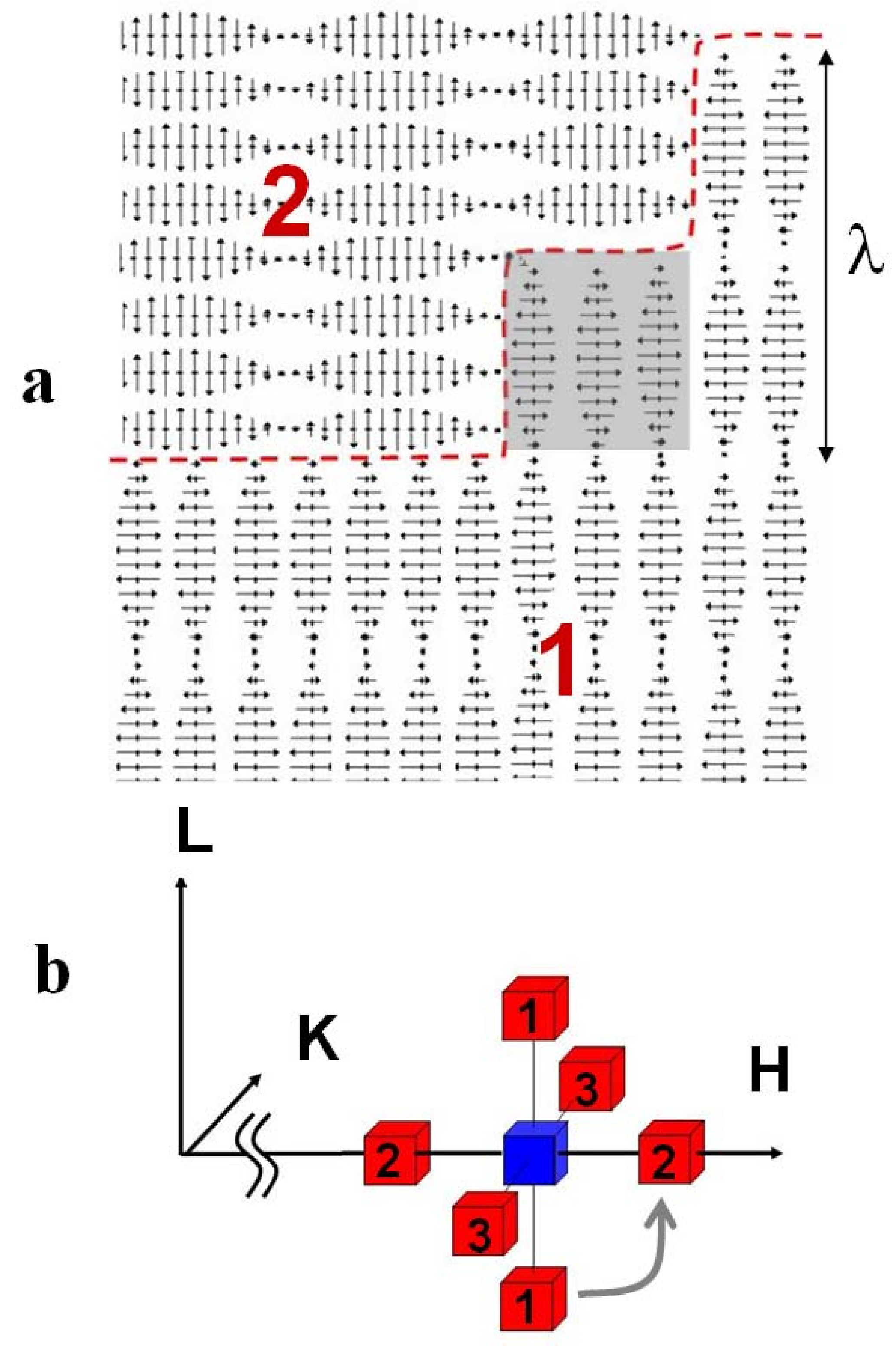}
\caption{\textbf{$\vert $ Spin Density Wave Domain Wall in
Chromium. a}, Schematic representation of the domain wall (red dashed line)
separating two regions with perpendicular orientations of transverse spin
density wave. The domain wall is shown to propagate along the weak nodal
planes -- where local magnetization approaches zero. Shaded region
represents an elementary domain unit with volume ($\lambda $/2)$^{3}$ that
can be thought of as a magnetic quantum dot in a cubic lattice of a similar
quantum dots. \textbf{b}, Reciprocal space configuration of lattice (200)
Bragg peak (blue) and six surrounding charge density wave satellites (red).
Domains marked as 1 and 2 in \textbf{a }contribute to pairs of satellites
marked 1 and 2 in \textbf{b}, respectively. 90 degree rotation of SDW
propagation vector within the shaded elementary volume of domain 1 would
realign spins with domain 2, resulting in shift of domain wall and transfer
of scattering intensity from satellite pairs 1 to 2, marked with an arrow in
\textbf{b}.}
\end{figure}
\newpage

\begin{figure}
\vspace{5mm}
\includegraphics[angle=0,width=0.6\columnwidth]{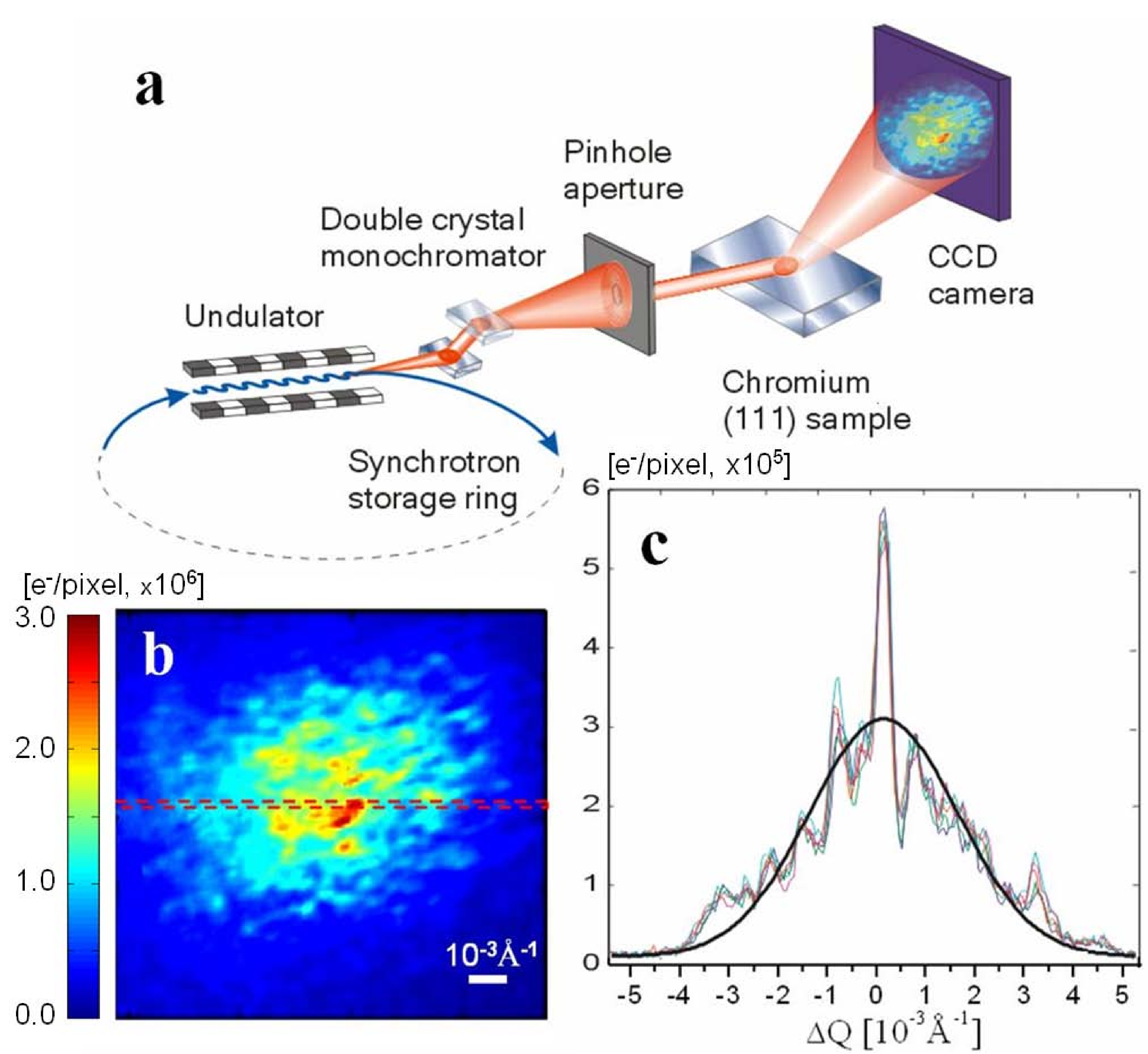}
\caption{\textbf{$\vert $ X-ray speckle measurements. a}, Schematic
of the experimental setup.\textbf{ b}, CCD image of the x-ray speckle
observed for the [200] lattice Bragg reflection. \textbf{c},\textbf{
}Intensity distribution for a line scan across a region shown with a bar in
(B) panel. Five differently coloured and nearly identical lines represent
line scans of the portion of speckle pattern shown with red dashed line in
\textbf{b}, taken one hour apart. Black line is a simulated statistically
averaged Gaussian profile, expected for completely incoherent beam.
}
\end{figure}
\newpage

\begin{figure}
\vspace{5mm}
\includegraphics[angle=0,width=0.8\columnwidth]{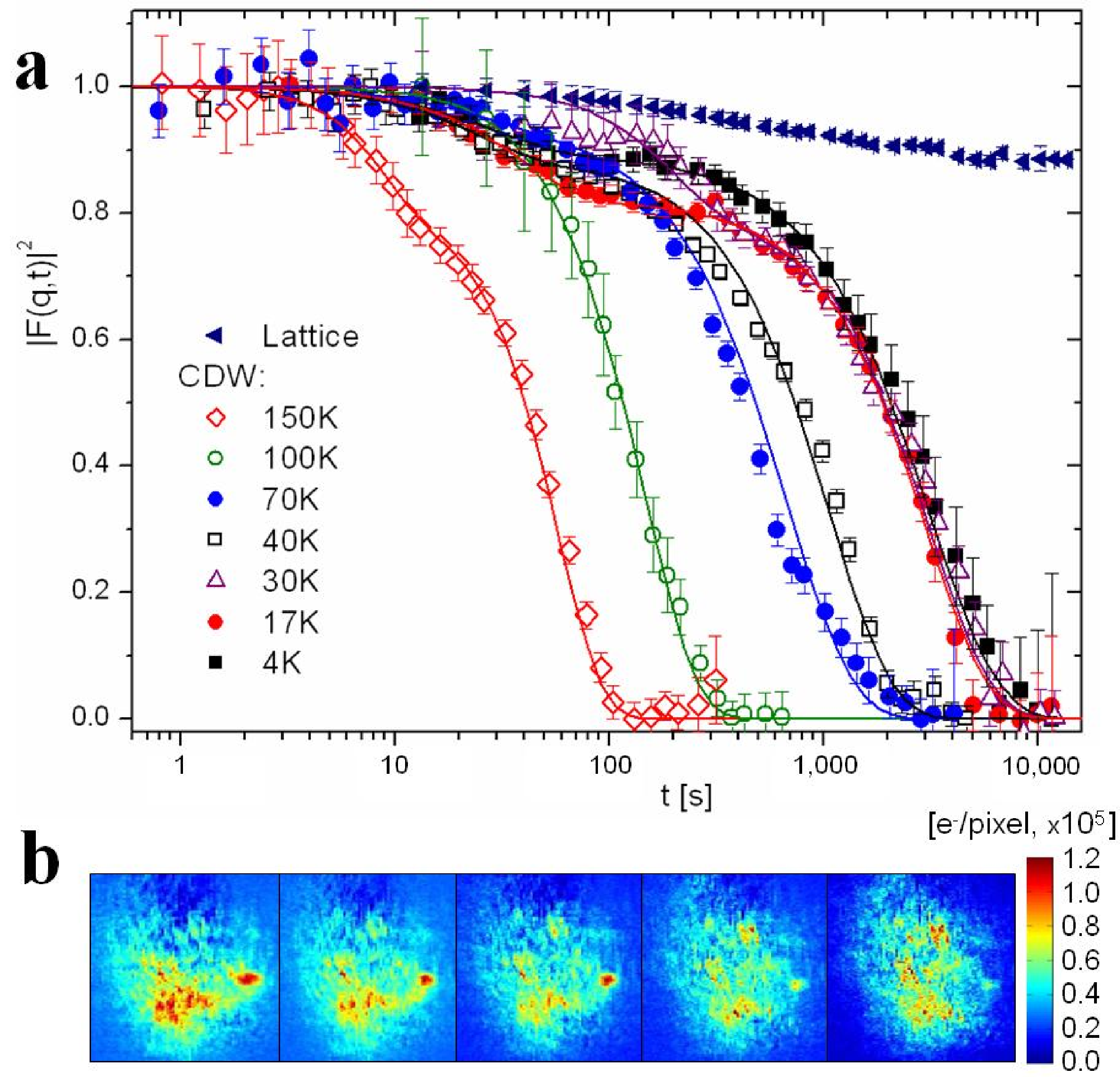}
\caption{\textbf{$\vert $ Autocorrelation of speckle images.}
\textbf{a,} Intensity autocorrelation data for [200] lattice Bragg peak as
well as for CDW superlattice [2-2$\delta $, 0, 0] peak at T=150 K, 100 K, 70
K, 40 K, 30 K, 17 K and 4 K. Two distinct timescales are clearly present in
the CDW autocorrelation function. Solid lines represent theoretical fits to
the data. See text for further details. \textbf{b,} Time sequence of CDW
speckle pattern evolution at 17 K. Subsequent images are taken 1,000 s
apart, each image is 10$^{ - 2}$ {\AA}$^{ - 1 }$by$^{ }$10$^{ - 2}$ {\AA}$^{
- 1}$.}
\end{figure}

\newpage
\begin{figure}
\vspace{5mm}
\includegraphics[angle=0,width=0.5\columnwidth]{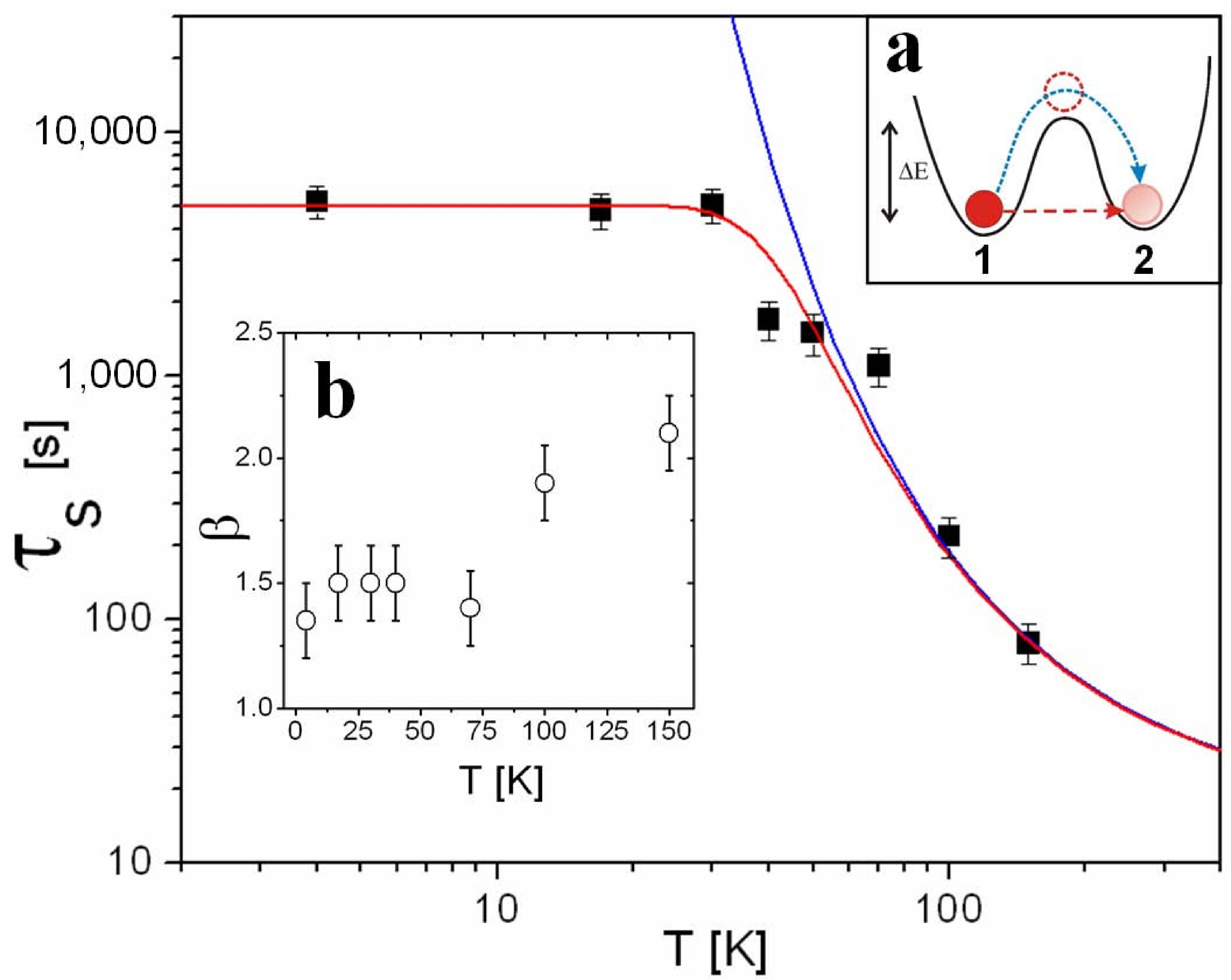}
\caption{\textbf{ $\vert $ Temperature-dependent domain wall
dynamics.}Characteristic slow fluctuation timescale $\tau _{s}$ obtained
from fits to autocorrelation function data shown in Fig. 3a, compared to
classical Arrhenius model (blue line) and a model that also includes a
temperature-independent switching rate term (red line). \textbf{a},
Potential energy surface including thermally activated (blue dashed line)
and quantum tunnelling (red dashed line) mechanisms of the transition
between the two low energy domain configurations 1 and 2 (see Fig. 1 for an
example involving elementary switching volume) separated by energy barrier
$\Delta $E \textbf{b}, Values of stretching exponent $\beta $ for various
temperatures.
}
\end{figure}

\end{document}